\documentclass[preprint2]{aastex}

\shorttitle{RHESSI CDS}

\shortauthors{Milligan et al.}

\begin{document}

\title{RHESSI and SOHO/CDS Observations of Explosive Chromospheric Evaporation}

\notetoeditor{the contact email is r.milligan@qub.ac.uk and is the only one which should appear on the journal version}

\author{Ryan O. Milligan\altaffilmark{1,3}, 
        Peter T. Gallagher\altaffilmark{2,3,4},
        Mihalis Mathioudakis\altaffilmark{1},
	D. Shaun Bloomfield\altaffilmark{1},
        Francis P. Keenan\altaffilmark{1}, and
	Richard A. Schwartz\altaffilmark{3,5}}

\altaffiltext{1} {Department of Physics and Astronomy, Queen's University Belfast, Belfast, BT7 1NN, Northern Ireland.}
\altaffiltext{2} {School of Physics, Trinity College Dublin, Dublin 2, Ireland.}
\altaffiltext{3} {Laboratory for Astronomy and Solar Physics, NASA Goddard Space Flight Center, Greenbelt, MD 20771, U.S.A.}
\altaffiltext{4} {L-3 Communications GSI.}
\altaffiltext{5} {Science Systems and Applications, Inc.}

\begin{abstract} 

Simultaneous observations of explosive chromospheric evaporation are presented
using data from the {\it Reuven Ramaty High Energy Solar Spectroscopic Imager
(RHESSI)} and the Coronal Diagnostic Spectrometer (CDS) onboard {\it SOHO}. For
the first time, co-spatial imaging and spectroscopy have been used to observe
explosive evaporation within a hard X-ray emitting region. {\it RHESSI} X-ray
images and spectra were used to determine the flux of non-thermal electrons
accelerated during the impulsive phase of an M2.2 flare. Assuming a thick-target
model, the injected electron spectrum was found to have a spectral index of
$\sim$7.3, a low energy cut-off of $\sim$20~keV, and a resulting flux of
$\geq$4$\times$10$^{10}$~ergs~cm$^{-2}$~s$^{-1}$. The dynamic response of the
atmosphere was determined using CDS spectra, finding a mean upflow velocity of
230$\pm$38~km~s$^{-1}$ in \ion{Fe}{19}~(592.23~\AA), and associated downflows of
36$\pm$16~km~s$^{-1}$ and 43$\pm$22~km~s$^{-1}$ at chromospheric and transition
region temperatures, respectively, relative to an averaged quiet-Sun spectra.
The errors represent a 1$\sigma$ dispersion. The properties of the accelerated
electron spectrum and the corresponding evaporative velocities were found to be
consistent with the predictions of theory.

\end{abstract}


\section{INTRODUCTION} 
\label{intro} 

Current solar flare models (\citealt{anti78};
\citealt{fish84,fish85a,fish85b,fish85c}; \citealt{mari89}) predict two types of
chromospheric evaporation processes. ``Gentle'' evaporation occurs when the
chromosphere is heated either directly by non-thermal electrons, or indirectly
by thermal conduction. The chromospheric plasma subsequently loses energy via a
combination of radiation and low-velocity hydrodynamic expansion. ``Explosive''
evaporation takes place when the chromosphere is unable to radiate energy at a
sufficent rate and consequently expands at high velocities into the overlying
flare loops. The overpressure of evaporated material also drives low-velocity
downward motions into the underlying chromosphere, in a process known as
chromospheric condensation. 

From a theoretical perspective, \cite{fish85a} investigated the relationship
between the flux of non-thermal electrons ($F$) and the velocity response of the
atmosphere for the two classes of evaporation. For gentle evaporation,
non-thermal electron fluxes of $\leq$ 10$^{10}$~ergs~cm$^{-2}$~s$^{-1}$ were
found to produce upflow velocities of tens of kilometres per second. In
contrast, explosive evaporation was found to be associated with higher
non-thermal electron fluxes ($F \geq$
3~$\times$~10$^{10}$~ergs~cm$^{-2}$~s$^{-1}$) which drive both upflows of hot
material at velocities of several hundred kilometres per second {\it and}
downflows of cooler material at tens of kilometres per second. 

\begin{figure}[!t]
\begin{center}
\includegraphics[width=8.0cm]{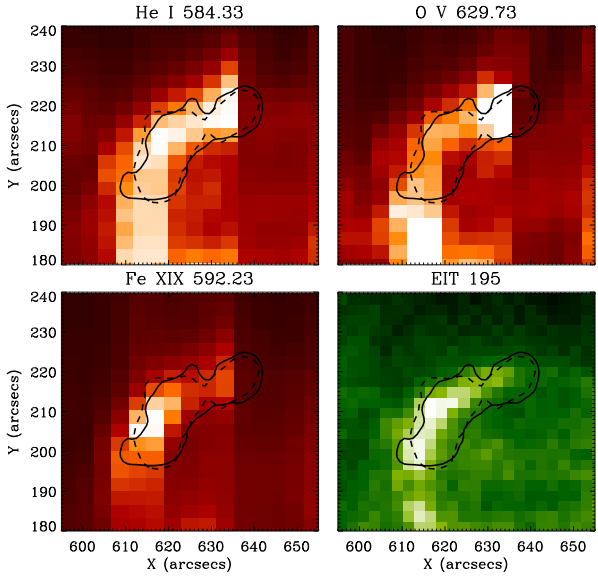}

\caption{CDS images in the \ion{He}{1}, \ion{O}{5}, and \ion{Fe}{19} emission
lines observed during the implusive phase of the flare, with the corresponding
EIT 195~\AA~image. {\it RHESSI} 12--25~keV ({\it dashed}) and 25--60~keV ({\it
solid}) contours are overlayed, each drawn at 10\% of the peak intensity.}

\label{hsi_cds_eit}
\end{center}
\end{figure}

Observationally, previous studies have identified blue-shifted Soft X-Ray and
EUV lines indicative of chromospheric evaporation. Using the Bent Crystal
Spectrometer onboard the {\it Solar Maximum Mission}, \cite{anto83} and
\cite{zarr88} reported upflow velocities of 400~km~s$^{-1}$ and 350~km~s$^{-1}$,
respectively, in \ion{Ca}{19} lines (3.1--3.2~\AA). More recently,
\cite{czay99}, \cite{teri03}, and \cite{delz05b} observed velocities of
140--200~km~s$^{-1}$ in \ion{Fe}{19} (592.23~\AA), using the Coronal Diagnostic
Spectrometer (CDS; \citealt{harr95}) onboard the {\it Solar and Heliospheric
Observatory (SOHO)}. Simultaneous upflows and downflows during a Hard X-Ray
(HXR) burst indicative of explosive evaporation have been observed using CDS and
{\it Yohkoh}/Hard X-Ray Telescope by \cite{bros04}. While these studies provided
a measurement of the dynamic response of the flaring chromosphere, they were
unable to provide a measurement of the flux of electrons responsible for driving
such motions, nor the spatial relationship between the two.

In this {\it Letter}, simultaneous {\it Reuven Ramaty High Energy Solar
Spectroscopic Imager} ({\it RHESSI}; \citealt{lin02}) and CDS observations are
combined for the first time to investigate the relationship between the
non-thermal electron flux and the response of the solar atmosphere. In
Section~\ref{obs} the analysis techniques employed are described, while the
results are presented in Section~\ref{results}. Our Conclusions are then given
in Section~\ref{disc}.

\section{OBSERVATIONS AND DATA ANALYSIS}
\label{obs}

This study focuses on a {\it GOES} M2.2 flare, which began at 12:44~UT on
2003 June 10. The event was selected from a sample of approximately 50 flares
jointly observed by {\it RHESSI} and CDS. The limited field of view, cadence,
and operating schedule of CDS, coupled with {\it RHESSI} nighttime and South
Atlantic Anomaly passes, make simultaneous observations by the two instruments
quite rare.

\subsection{The Coronal Diagnostic Spectrometer (CDS)}
\label{cds}

The CDS observations reported here were obtained with the {\it FLARE\_AR}
observing sequence. {\it FLARE\_AR} contains five $\lesssim$4~\AA\ wide spectral
windows centered on \ion{He}{1} (584.33~\AA; log T = 4.5), \ion{O}{5}
(629.73~\AA; log T = 5.4), \ion{Mg}{10} (624.94~\AA; log T = 6.1), \ion{Fe}{16}
(360.76~\AA; log T = 6.4), and \ion{Fe}{19} (592.23~\AA; log T = 6.9). Each
raster consists of 45 slit positions, each $\sim$15 seconds long, resulting in
an effective cadence of $\sim$11 minutes. The slit itself is
4\arcsec$\times$180\arcsec~yielding a $\sim$180\arcsec$\times$180\arcsec~field
of view. A zoomed-in region of the \ion{He}{1}, \ion{O}{5}, and \ion{Fe}{19}
rasters from the impulsive phase of the flare is given in
Figure~\ref{hsi_cds_eit}. Also shown is the Extreme ultraviolet Imaging
Telescope (EIT; \citealt{dela95}) 195~\AA\ passband image obtained at 12:48~UT.
A series of subsequent EIT images makes it clear that the He I and O V brightenings come from a flare ribbon rather than two distinct footpoints as Figure~\ref{hsi_cds_eit} may suggest.

\begin{figure*}[!t]
\begin{center}
\includegraphics[height=16.5cm,angle=90]{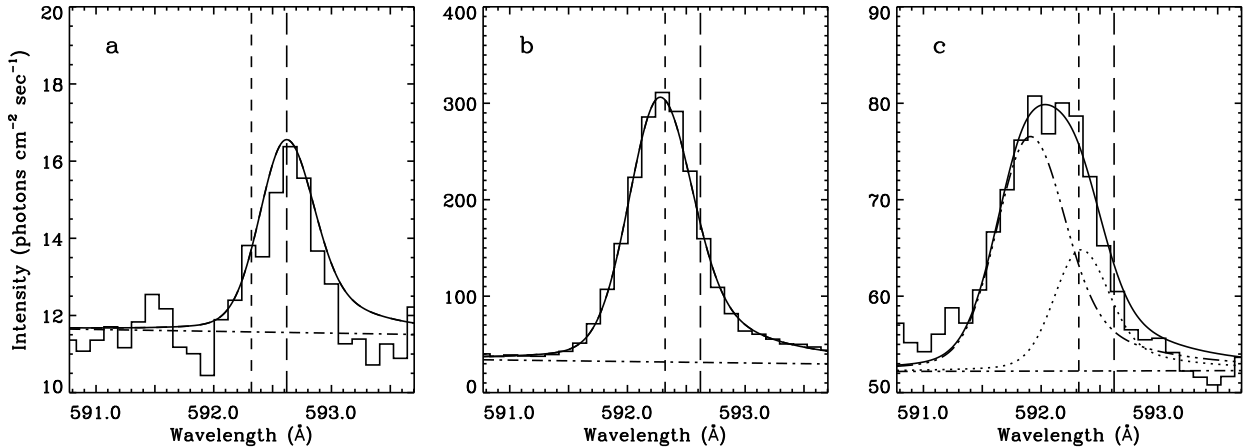}

\caption{Sample of spectra from the spectral window centered on
\ion{Fe}{19} (592.23~\AA). The vertical {\it short-dashed} line shows the rest
wavelength of the \ion{Fe}{19} line at 592.32~\AA~while the {\it long-dashed}
line shows the rest wavelength of the \ion{Fe}{12} line at 592.62~\AA. Panel `a'
was obtained from a quite Sun region, panel `b' shows a stationary \ion{Fe}{19}
line from a post-implusive phase flare kernel, while panel `c' shows an
\ion{Fe}{19} line from the flare ribbon during the implusive phase. The {\it
dotted} line indicates the stationary component while the {\it
triple-dot-dashed} line indicates the blue-shifted component.}

\label{cds_spec}
\end{center}
\end{figure*}

The spectrum from each CDS pixel was fitted with a broadened Gaussian profile
\citep{thom99}, for each of the five spectral windows.  Velocity maps were
created by measuring Doppler shifts relative to quiet-Sun spectra, which were
assumed to be emitted by stationary plasma. Preliminary fits to the \ion{Fe}{19}
line during the impulsive phase of the flare revealed an asymmetric broadening
beyond the instrumental resolution of CDS. The strongest blue asymmetries were
found within the flare ribbon during the impulsive phase.  Outside this area,
and after the impulsive phase, the Fe XIX line was observed to have a width
comparible to the instrumental width. Figure~\ref{cds_spec} shows a sample of
spectra taken from the spectral window centered on the \ion{Fe}{19} (592.23~\AA)
emission line. Panel `a' shows a spectrum from a quiet Sun area in which no
\ion{Fe}{19} emission was visible. Instead, a weak emission line was observed at
592.6~\AA~which \cite{delz05a} have identified as \ion{Fe}{12}. Panel `b' shows
a stationary \ion{Fe}{19} emission line extracted from a bright region, but
after the impulsive phase at $\sim$12:50~UT when no significant flows are
expected. An emission line with a strong blue asymmetry is shown in panel `c'.
This was extracted from the flare ribbon during the impulsive phase. The best
fit to this line was consistent with stationary and blue-shifted components,
both with widths comparible to the instrumental resolution. As \ion{Fe}{19} is
not observed in quiet-Sun spectra, and following from \citet{teri03}, the
Doppler velocity was measured as the shift between these two components. A
heliographic correction was also applied, due to the longitude of the
observations and assuming purely radial flows, 

\begin{figure}[!t] 
\begin{center} 
\includegraphics[width=8.0cm]{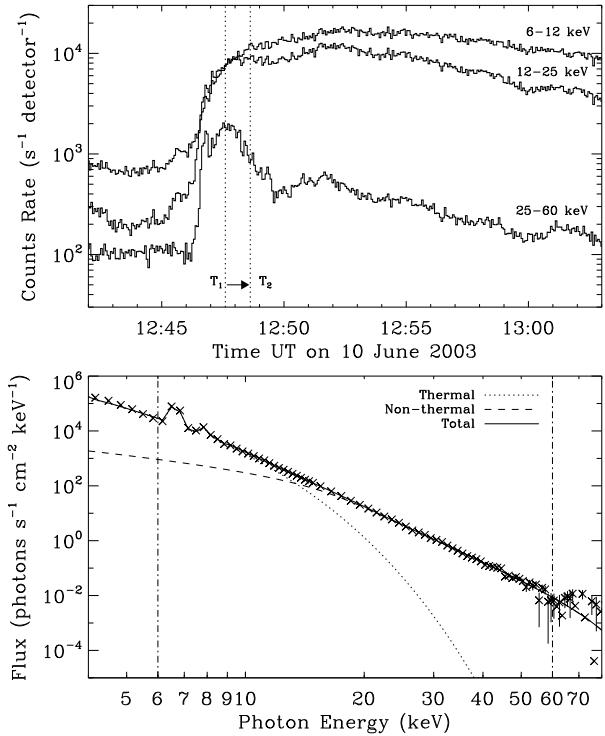}

\caption{{\it Top panel:} {\it RHESSI} lightcurves from the 6--12, 12--25, and
25--60~keV bands. The {\it dotted} vertical lines indicate the time interval
over which images and spectra were obtained to correspond to time when
significant upflows were observed using CDS. {\it Bottom panel:} Portion of the
{\it RHESSI} spectrum integrated over the timerange given above. The energy
range 6--60~keV (vertical {\it dot-dash} lines) was fitted with an isothermal
component ({\it dotted curve}) and a thick-target bremsstrahlung component ({\it
dashed curve}).}

\label{hsi_spec} 
\end{center}
\end{figure}

\subsection{The Reuven Ramaty High Energy Spectroscopic Imager (RHESSI)}
\label{rhessi}

{\it RHESSI} is an imaging spectrometer capable of observing X- and $\gamma$-ray
emission over a wide range of energies ($\sim$3~keV--17~MeV). During this event
the thin attenuators on {\it RHESSI} were in place thus limiting the energy
range to $\gtrsim$6~keV. Flare emission was not observed above $\sim$60~keV. The
flare lightcurves are shown in the top panel of Figure~\ref{hsi_spec}. Both the
{\it RHESSI} images and spectra were obtained over a 64 second period from
12:47:34--12:48:38~UT to coincide with the timerange over which CDS observed
blue asymmetries in the \ion{Fe}{19} line. This time interval lies within the
impulsive 25--60~keV HXR burst and is indicated by two vertical dotted lines in
the top panel of Figure~\ref{hsi_spec}. {\it RHESSI} images in two energy bands
(12--25 and 25--60~keV) were reconstructed using the {\it Pixon} algorithm
\citep{hurf02}. Contours at 10\% of the peak intensity in each band are
overlayed on each EUV image in Figure~\ref{hsi_cds_eit}. 

The {\it RHESSI} spectrum was fitted assuming an isothermal distribution at low
energies, and thick-target emission at higher energies (bottom panel of
Figure~\ref{hsi_spec}). A thick-target model was chosen over a thin-target model
as it is believed that the density of the flare loop is insufficent to
thermalise the electrons as they propagate to the chromosphere. The thick-target
model is used in the vast majority of cases (e.g. \citealt{holm03},
\citealt{vero05}). Furthermore, in this flare the HXR source is clearly aligned
with the He I ribbon as seen by CDS, which implies that the accelerated
electrons are losing their energy in the dense chromosphere rather than in the
coronal loops. The total power of non-thermal electrons above the low energy
cut-off ($\epsilon_{c}$) was calculated from $P(\epsilon \geq \epsilon_{c}) =
\int_{\epsilon_{c}}^{\infty} f_{e}(\epsilon)d\epsilon$ ergs~s$^{-1}$, where 
$f_{e}(\epsilon)$ $\sim$ $\epsilon^{-\delta}$ electrons keV$^{-1}$ s$^{-1}$ is
the thick-target electron injection spectrum and $\delta$ is the associated
spectral index \citep{brow71}. Because of the steepness of the {\it RHESSI}
spectrum at high energies, the non-thermal flux is quite sensitive to the value
of the low energy cut-off.  In order to put a constraint on this value, the
temperature of the thermal component was obtained by another independent method,
i.e. the equivalent width of the Fe line complex at 6.7 keV (\citealt{phil04}).
The value of the equivalent width of this line, which is quite sensitive to the
temperature, was used to estimate the temperature of the thermal component.
Having fixed this value, the entire {\it RHESSI} spectrum was fitted using a
least-squares fit. 

\section{RESULTS}
\label{results}

The thick-target model fitted to the {\it RHESSI} spectrum in
Figure~\ref{hsi_spec} was consistent with an electron distribution having
$\epsilon_c\sim$20~keV and $\delta\sim$7.3. The break energy of 20~keV is
consistent with earlier works (e.g. \citealt{holm03}, \citealt{sui05}). The
total power in non-thermal electrons was therefore
1$\times$10$^{29}$~ergs~s$^{-1}$. Exploring the possible range of values for the
break energy for this flare would yield an electron power value of
4$\times$10$^{29}$~ergs~s$^{-1}$ for $\epsilon_c$ = 15.0~keV while $\epsilon_c$
= 25.0~keV would give a power value of 6$\times$10$^{28}$~ergs~s$^{-1}$.
However, either of these break energies would give a worse $\chi^{2}$ value than
obtained from the original fit. By comparison, the total thermal power for the
same time interval was found to be 1.2$\times$10$^{28}$~ergs~s$^{-1}$. 

Using the reconstructed 25--60~keV image, an upper-limit to source size was
calculated to be 2.3$\times$10$^{18}$~cm$^{2}$. This was found by summing over
all pixels with counts greater than 10\% of the peak value. This threshold was
chosen to eliminate sources outside of the main HXR-emitting region, which were
assumed to be unreal; the source area was not found to be highly sensitive to
this value. For example, a threshold of 5\% yielded an area of
3.2$\times$10$^{18}$~cm$^{2}$ and 15\% yielded 2$\times$10$^{18}$~cm$^{2}$.
This area was also confirmed using the Fourier modulation profiles from each of
{\it RHESSI}'s nine detectors, which are sensitive to spatial scales from
2.2\arcsec to 183\arcsec. Assuming a filling factor of unity, the resulting
non-thermal electron flux was calculated to be
$\geq$4$\times$10$^{10}$~ergs~cm$^{-2}$~s$^{-1}$. 

\begin{figure}[!t]
\begin{center}
\includegraphics[width=8.0cm]{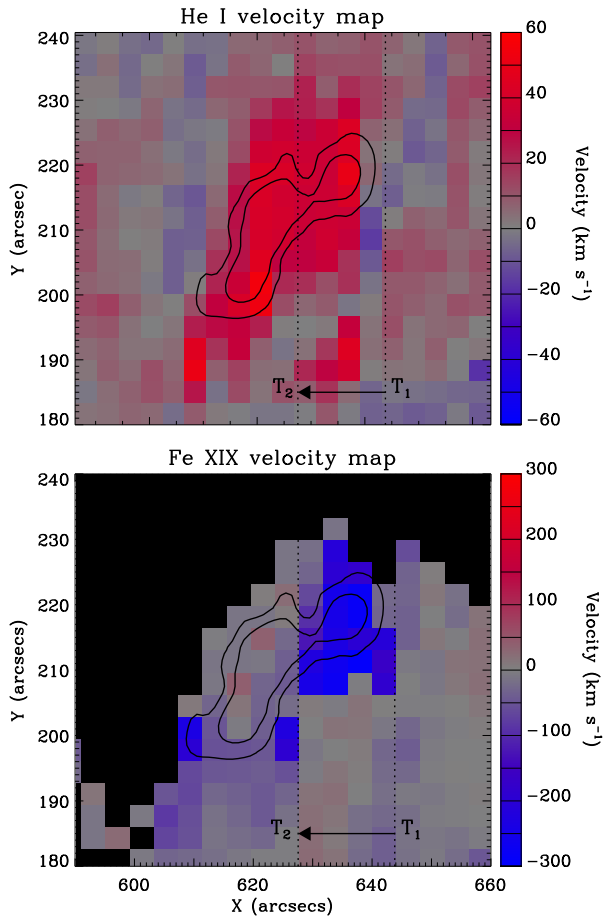}

\caption{Velocity maps in \ion{He}{1} and \ion{Fe}{19}. Downflows are indicated
by red pixels while upflows are indicated by blue pixels. The vertical {\it
dashed} lines correspond to the times indicated by the vertical {\it dashed}
lines in Figure~\ref{hsi_spec} and the arrow denotes the direction in which the
CDS slit moves. Black regions in the \ion{Fe}{19} map represent pixels where no
significant \ion{Fe}{19} emission was observed. {\it RHESSI} 25--60~keV contours
at 10\% and 40\% of the peak intensity are overlayed.}

\label{vel_maps}
\end{center}
\end{figure}

\begin{figure}[!t]
\begin{center}
\includegraphics[height=8.0cm,angle=90]{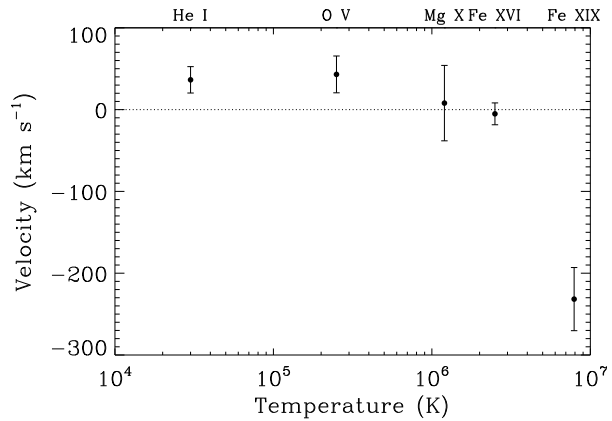}

\caption{Plasma velocity as a function of temperature for the five lines
observed using CDS. Positive velocities indicate downflows, while negative
values indicate upflows.}

\label{vel_plot}
\end{center}
\end{figure}

Figure~\ref{vel_maps} shows velocity maps in the \ion{He}{1} and \ion{Fe}{19}
lines. The \ion{He}{1} map shows consistent downflows of 20--50~km~s$^{-1}$
until the slit leaves the flaring region at $\sim$12:50~UT. A velocity map in
\ion{O}{5} showed a similar trend. However, the \ion{Fe}{19} map shows strong
upflow velocities of 190--280~km~s$^{-1}$ during the HXR peak, indicated by
`T$_{1}$' and `T$_{2}$' on Figures~\ref{hsi_spec} and \ref{vel_maps}. No
significant upflows were evident once the HXRs begin to diminish from time
T$_{2}$ onwards.

By identifying each pixel in the \ion{Fe}{19} map that required a two-component
fit to the line profile, between times T$_{1}$ and T$_{2}$, the velocity was
measured for the corresponding pixel in each of the five CDS rasters.
Figure~\ref{vel_plot} shows the mean velocity as a function of temperature for
each line using the methods described in Section~\ref{obs}. The error bars
represent a 1$\sigma$ dispersion. At chromospheric and transition region
temperatures, plasma velocities show red-shifts of 36$\pm$16~km~s$^{-1}$ and
43$\pm$22~km~s$^{-1}$, respectively, while the blue-shift observed in the 8~MK
\ion{Fe}{19} line corresponds to a velocity of 230$\pm$38~km~s$^{-1}$. No
significant flows were observed in the \ion{Mg}{10} and \ion{Fe}{16} lines. The
combination of high-velocity upflows and low-velocity downflows, together with a
non-thermal electron flux of $\geq$4$\times$10$^{10}$~ergs~cm$^{-2}$~s$^{-1}$
provides clear evidence for explosive chromospheric evaporation.

\section{DISCUSSION AND CONCLUSIONS}
\label{disc}

For the first time, co-spatial and co-temporal HXR and EUV observations of
chromospheric evaporation are presented using {\it RHESSI} and {\it SOHO}/CDS.
High upflows velocities ($\sim$230~km~s$^{-1}$) were clearly observed in
high-temperature \ion{Fe}{19} emission during the impulsive phase of an M2.2
flare, while much lower downflow velocities ($\sim$40~km~s$^{-1}$) were observed
in the cooler \ion{He}{1} and \ion{O}{5} lines. The value of the non-thermal
electron flux ($\geq$4$\times$10$^{10}$~ergs~cm$^{-2}$~s$^{-1}$) and the
resulting velocity response are indicative of an explosive evaporation process
occuring during this flare, as laid out in \cite{fish85a} and \cite{mari89}.

The combination of HXR and EUV observations presented in this {\it Letter} have
enabled us to obtain a greater understanding of the characteristics of
chromospheric evaporation, a fundamental process in solar flares. We have
presented the first detection of explosive mass motions within HXR footpoints,
and determined the flux of non-thermal electrons responsible for driving such
flows. 

\acknowledgments  

This work has been supported by a Department of Employment and Learning (DEL)
studentship and a Cooperative Award in Science and Technology (CAST). FPK is
grateful to AWE Aldermaston for the award of a William Penny Fellowship. We
would like to thank Drs. Brian Dennis, Joe Gurman, and Dominic Zarro at NASA
Goddard Space Flight Center for their stimulating discussion and continued
support. We would also like to thank the anonymous referee for their comments
and suggestions which have greatly improved this {\it Letter}. SOHO is a project
of international collaboration between the European Space Agency (ESA) and NASA.

\end{document}